\documentclass[prb,reprint,a4paper,showpacs,citeautoscript,floatfix]{revtex4-1}
\pdfoutput=1
\usepackage[charter]{mathdesign}
\usepackage{amsmath}
\usepackage[pdftex]{graphicx}
\usepackage{microtype}
\usepackage{bm}\let\vec\bm

\usepackage[svgnames]{xcolor}
\usepackage[
	colorlinks=True,linkcolor=DarkRed,citecolor=ForestGreen,urlcolor=MediumBlue,
	pdfstartview=FitH,bookmarks=False,pdfpagemode=UseNone
]{hyperref}

\begin{document}

\title{Quasiparticle spectra of Abrikosov vortices in a uniform supercurrent flow}

\author{C. Berthod}
\affiliation{DPMC, University of Geneva, 24 quai Ernest-Ansermet, 1211 Geneva 4,
Switzerland}

\date{September 15, 2013}

\begin{abstract}

We calculate the local density of states of a vortex in a two-dimensional
$s$-wave superconductor, in the presence of a uniform applied supercurrent. The
supercurrent induces changes in the electronic structure for the isolated vortex
as well as the vortex lattice, which agree with the recent measurements in
$2H$-NbSe$_2$ [Maldonado \textit{et al}.,
\href{http://dx.doi.org/10.1103/PhysRevB.88.064518}{Phys. Rev. B \textbf{88},
064518 (2013)}]. We find that the supercurrent polarizes the core states when
the vortices are pinned. This shows that the transfer of momentum from the
supercurrent to the bound states and the rigidity of the wave functions must be
considered for understanding the various forces acting on collectively pinned
Abrikosov vortices.

\end{abstract}

\pacs{74.25.N-, 74.55.+v, 74.81.-g}
\maketitle

The quantum states bound to magnetic vortices in type-II superconductors carry
information about the fundamental properties of the superconducting state. The
existence of bound states was predicted long ago \cite{Caroli-1964}, but the
direct observation in NbSe$_2$ awaited the invention of the scanning tunneling
microscope (STM) \cite{Hess-1989}. The complete mapping of the tunneling
conductance in real space as a function of applied bias provided a large data
set, in striking agreement with the BCS predictions for the local density of
states (LDOS) of a vortex \cite{Gygi-1990a, *Gygi-1991}. Since then, several
groups have investigated the vortex cores by STM in NbSe$_2$ \cite{Renner-1991,
Behler-1994, Troyanovski-1999}, in other classical superconductors
\cite{DeWilde-1997b, Sakata-2000, Eskildsen-2002, Sosolik-2003,
Guillamon-2008a}, in high-$T_c$ cuprates, \cite{[{See, e.g., }][{, and
references therein.}]Fischer-2007} and, more recently, in the pnictides
\cite{Yin-2009, Shan-2011, Song-2011, Hanaguri-2012}. While in classical
superconductors, including the pnictides, these studies usually reveal
vortex-core spectra in good qualitative agreement with the BCS theory,
significant deviations are found in high-$T_c$ superconductors, probably due to
an anomalous normal state \cite{Fischer-2007}. The interpretation of vortex-core
spectra in the cuprates remains an open question.

Recently, a measurement of the vortex electronic structure in the presence of an
in-plane current flow was performed in NbSe$_2$ \cite{Maldonado-2013}. When a
supercurrent is established across a vortex lattice, a ``Lorentz force'' acts on
the vortices in the direction normal to the applied current
\cite{deGennes-1999}. In Ref.~\onlinecite{Maldonado-2013}, the current was
sufficiently small for the force to remain below the depinning threshold, and
the authors could map the LDOS of static vortices with and without the applied
current. The main observation of this experiment is that the application of a
current transfers low-energy spectral weight from inside the cores, where the
zero-bias conductance is reduced, to in between the vortices where it is
enhanced, while the converse appears at the gap edges, where the spectral weight
is enhanced inside the cores and depleted outside. The measurements also suggest
that the current increases the size of the vortex cores. To interpret these
trends, the authors assume that the applied current reduces the smallest gap on
the two-band Fermi surface of NbSe$_2$. This would affect the formation of
Andreev bound states in the cores, diminishing their energy separation.

This interpretation refers to second-order changes in the modulus of the order
parameter but ignores that the leading effect of the applied current is a
distortion of the order-parameter \emph{phase}. From a mesoscopic point of view,
a uniform supercurrent in a vortex lattice can be regarded as a distortion of
the phase. The modulus of the pair wave function
$\Psi(\vec{r})=|\Psi(\vec{r})|e^{i\chi(\vec{r})}$ vanishes at the vortex centers
and approaches the constant zero-field value at a distance $r_c\approx\xi$ from
the cores, where $\xi$ is the superconducting coherence length. Its phase
$\chi(\vec{r})$ winds by $2\pi$ around each vortex. The topological defect
associated with the phase winding is responsible for the formation of the vortex
bound states \cite{Volovik-1993a, Berthod-2005}. The supercurrent
$\vec{J}_{\chi}\approx(e\hbar/m)|\Psi|^2\vec{\nabla}\chi$ (neglecting magnetic
contributions) circulates around each vortex. Its intensity vanishes linearly in
the cores, decreases as $1/r$ at intermediate distances shorter than the
penetration depth, and is maximum at a distance $\sim r_c$ from the core
centers. In the presence of an applied uniform superflow, the pair wave function
becomes $\Psi(\vec{r})=|\Psi(\vec{r})|e^{i[\chi(\vec{r})+\vec{q}\cdot\vec{r}]}$,
where the applied current $\vec{J}_q\approx(e\hbar/m)|\Psi|^2\vec{q}$ vanishes
in the vortex cores like the vortex-lattice supercurrent. The phase distortion
displaces the electronic levels by the Doppler shift effect \cite{Volovik-1993},
and is therefore expected to change the vortex LDOS.

The effect of this phase distortion on the LDOS is studied here in a simple
one-band tight-binding model in two dimensions. This is not intended to be a
realistic model for NbSe$_2$. However, the features demonstrated here are
expected to be generic, and to apply to more sophisticated models as well. The
tight-binding and superconducting parameters are chosen in a way that allows a
semiquantitative comparison with NbSe$_2$. In a previous study, it was shown
that the vortex-core LDOS is weakly sensitive to distortions of the phase which
are random but is qualitatively modified by distortions which carry a
topological defect, like a nearby antivortex \cite{Berthod-2005}. The case of a
uniform distortion was not considered. In the present study, we keep the modulus
of the order parameter fixed and perturb the phase in order to simulate a
uniform applied current. This produces an exchange of spectral weight between
the core and the outside and an apparent increase of the vortex-core size, both
for an isolated vortex and for a vortex lattice. All trends observed in the
NbSe$_2$ experiment \cite{Maldonado-2013} can therefore be attributed to the
first-order effect of the applied current, without resorting to a reduction of
the order-parameter amplitude and/or to multiband effects.

The model is a tight-binding square lattice with a dispersion
$\xi_{\vec{k}}=-2t[\cos(k_xa)+\cos(k_ya)]-\mu$, and an $s$-wave superconducting
gap $\Delta$. We use $\Delta$ as the unit of energy, the lattice parameter $a$
as the unit of length, and we set the chemical potential to $\mu=2t>0$. This
locates the van Hove singularity at the positive energy $2t$, and produces an
electronlike Fermi surface corresponding to an electron density $n\approx0.4$
and a Fermi wave vector $k_{\mathrm{F}}\approx\pi/2$. With this choice, and if
$\Delta<t$, the normal-state DOS is approximately constant---reducing the
band-structure effects to a minimum---in the energy range $\pm3\Delta$, where we
aim to study the effect of the applied current. The isolated vortex and the
vortex-lattice structures are studied in a finite mesh of size $M\times M$
($M=71$). After computing the lattice Green's function,
	\begin{equation}
		G_{\vec{r}\vec{r}'}(\varepsilon)=G^0_{\vec{r}\vec{r}'}(\varepsilon)
		+\sum_{\vec{r}_1\vec{r}_2}G^0_{\vec{r}\vec{r}_1}(\varepsilon)
		\Sigma_{\vec{r}_1\vec{r}_2}(\varepsilon)G_{\vec{r}_2\vec{r}'}(\varepsilon),
	\end{equation}
the relation
$N(\vec{r},\varepsilon)=-(2/\pi)\text{Im}\,G_{\vec{r}\vec{r}}(\varepsilon)$
allows one to obtain the LDOS. The normal-state Green's function
$G^0_{\vec{r}\vec{r}'}(\varepsilon)=(1/N^2)
\sum_{\vec{k}}e^{i\vec{k}\cdot(\vec{r}-\vec{r}')}/
(\varepsilon-\xi_{\vec{k}}+i0^+)$ is calculated on a much larger $N\times N$
mesh ($N=1024$), taking advantage of the translation invariance. The self-energy
is \cite{Gorkov-1958}
	\begin{equation}\label{eq:Sigma}
		\Sigma_{\vec{r}\vec{r}'}(\varepsilon)=-\Psi(\vec{r})
		G^0_{\vec{r}'\vec{r}}(-\varepsilon)\Psi^*(\vec{r}'),
	\end{equation}
where the local $s$-wave order parameter $\Psi(\vec{r})$ describes an isolated
vortex at the central site or a vortex lattice, as well as the applied current.

\begin{figure}[b]
\includegraphics[width=0.7\columnwidth]{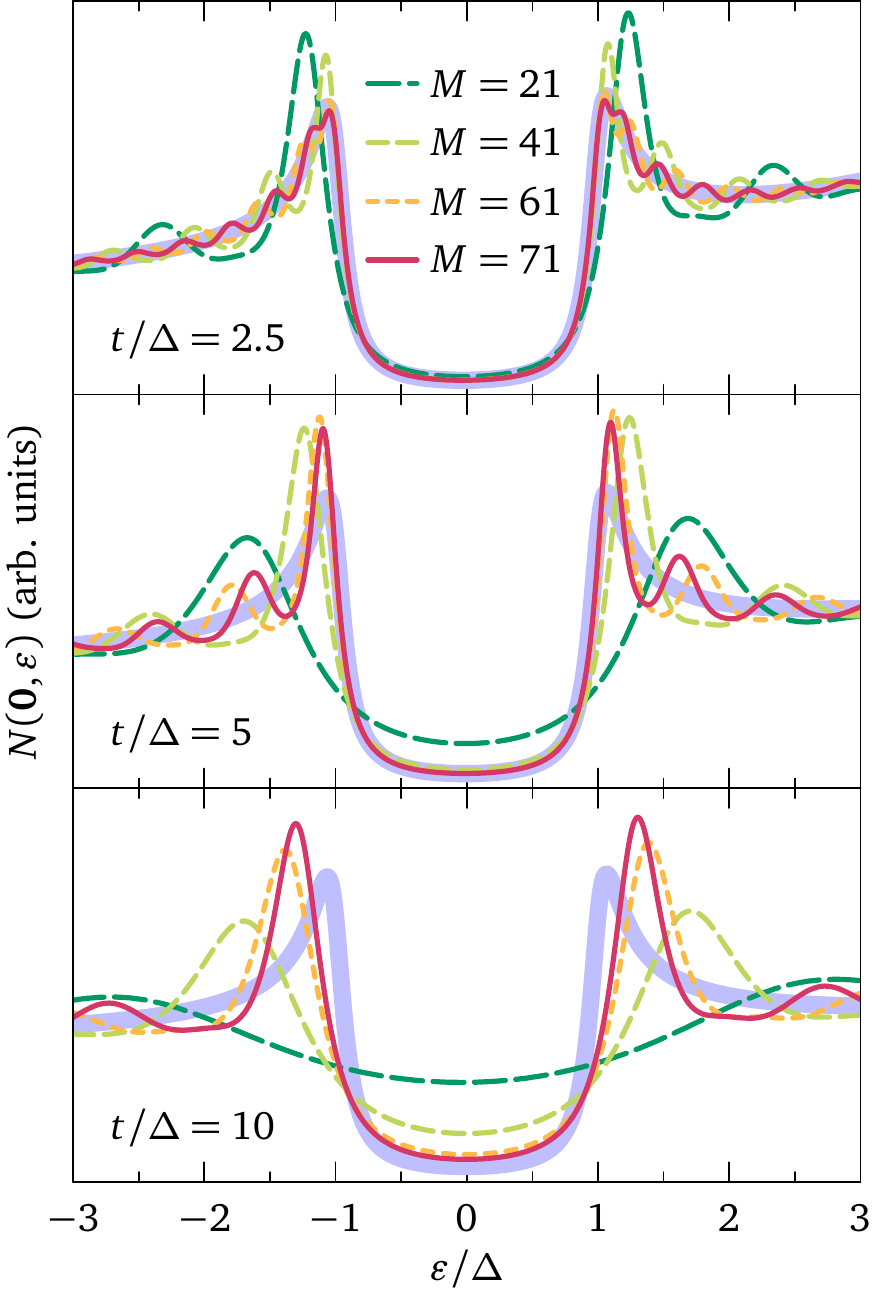}
\caption{\label{fig:fig1}
Finite-size effects. The thick solid lines show the exact DOS calculated for
each value of $t$ at zero field and without applied current. The thin lines show
the LDOS calculated at the central site of a $M\times M$ mesh, setting
$\Psi(\vec{r})=\Delta$ in Eq.~(\ref{eq:Sigma}). Finite-size effects are small
for $t=2.5$ and $M=71$ but remain significant for the largest $M$ if $t=10$,
even at subgap energies.
}
\end{figure}

In the experiment \cite{Maldonado-2013}, the zero-field spectrum is considerably
broadened as compared to an ideal $s$-wave superconducting DOS. This is not due
to the finite temperature, as the latter was set to 200~mK, which is $\sim100$
times smaller than the NbSe$_2$ gap of $\sim1.3$~meV. Impurity scattering is the
next candidate. We introduce a phenomenological impurity scattering through the
substitution $\varepsilon\to\varepsilon+i\Gamma$ in the definition of
$G^0_{\vec{r}\vec{r}'}(\varepsilon)$. Setting $\Gamma=0.1$, we obtain a
zero-field spectrum in good qualitative agreement with the NbSe$_2$ spectrum.
This value of $\Gamma$ will be used throughout. Before fixing the hopping $t$,
we need to consider finite-size effects. The latter are often overlooked in LDOS
calculations for vortices, but can be significant, even on a mesh as large as
$71\times71$. In our setup, the finite-size effects increase with increasing
$t$, as shown in Fig.~\ref{fig:fig1}. In order to have small finite-size effects
with $M=71$, we must take $t\lesssim2.5$. This is not far from the quantum limit
$k_{\mathrm{F}}\xi=1$. Using the BCS relation
$k_{\mathrm{F}}\xi=2E_{\mathrm{F}}/(\pi\Delta)$ and the value
$E_{\mathrm{F}}=2t$ corresponding to our dispersion, we estimate
$k_{\mathrm{F}}\xi=(4/\pi)(t/\Delta)\lesssim 3$. The typical value for NbSe$_2$
may be estimated as $k_{\mathrm{F}}\xi=m^*v_{\text{F}}\xi/\hbar\sim11$--$14$,
using the values $m^*=2m$, $v_{\text{F}}=8.2\times10^6\mathrm{cm/s}$, and
$\xi=79$--$100$~\AA\ reported in Ref.~\onlinecite{Gygi-1991}. The calculation
for higher values of $k_{\mathrm{F}}\xi$ require larger $M$, but the calculation
scales as $M^4$. One may however argue that the \emph{variations} of the LDOS
induced by the applied current are less sensitive to the boundary than the LDOS
itself. Hereafter we will present results for $t=2.5$ and for $t=10$
($k_{\mathrm{F}}\xi\sim 13$), focusing in the latter case, which is relevant for
a comparison with NbSe$_2$, on the LDOS variations induced by the applied
current.

\begin{figure}[tb]
\includegraphics[width=\columnwidth]{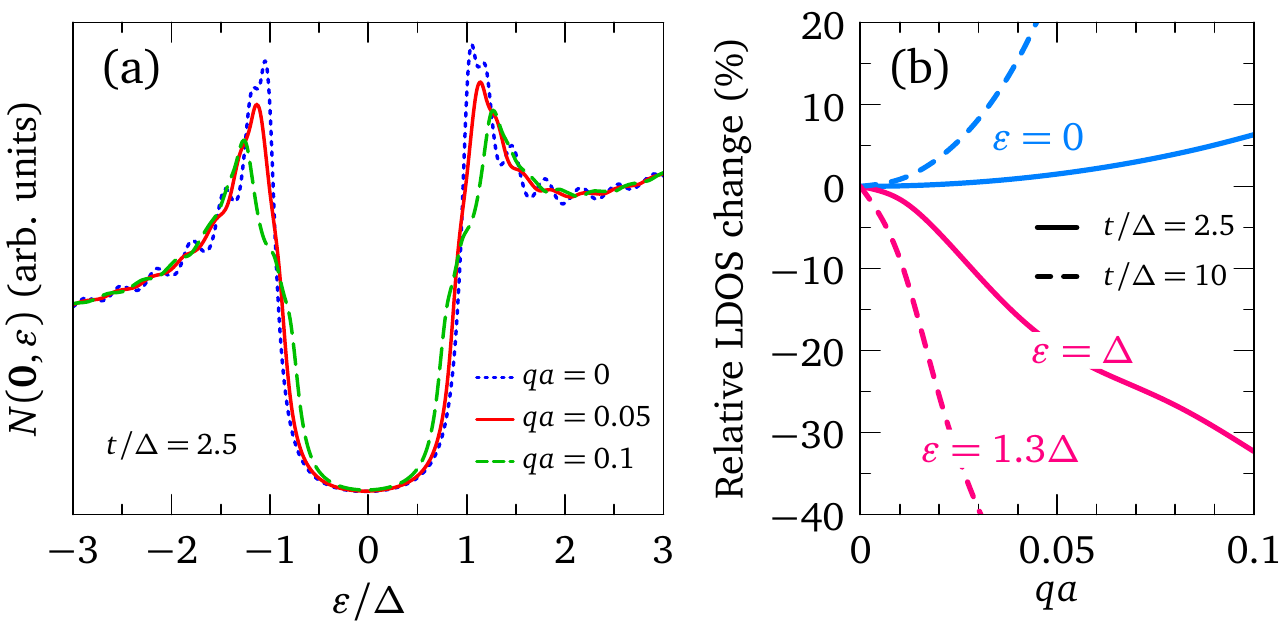}
\caption{\label{fig:fig2}
(a) Zero-field LDOS at the center of the $71\times71$ mesh in the absence of
superflow (dotted line), and in a uniform superflow with $q=0.05$ (solid line)
and $q=0.1$ (dashed line). (b) Relative variation of the LDOS at zero energy and
at the gap edge versus the applied current. The solid lines are for $t=2.5$
($k_{\mathrm{F}}\xi\sim 3$); the dashed lines are for $t=10$
($k_{\mathrm{F}}\xi\sim 13$). Due to finite-size effects, the gap edge is at
$1.3$ for $t=10$.
}
\end{figure}

The value of the current flowing around the vortices below the STM tip is not
known precisely \cite{Maldonado-2013}. Our approach to calibrate the current in
the model is to require that its effect on the zero-field LDOS is similar to the
observations made at zero field in NbSe$_2$. We consider a uniform superflow
along $x$ at zero field by setting $\Psi(\vec{r})=\Delta e^{-iqx}$ in
Eq.~(\ref{eq:Sigma}). The resulting LDOS is compared with the zero-current LDOS
in Fig.~\ref{fig:fig2}. The applied current increases the conductance in the gap
and decreases the conductance at the gap edges, consistent with the observations
\cite{Maldonado-2013}. We obtain a semiquantitative agreement with the
measurements performed in a current of 10.6~mA, namely, an $\sim20\%$ drop of
the conductance at the peak energy---by setting $q=0.05$ for $t=2.5$, and
$q=0.02$ for $t=10$. With these values, the applied current remains much smaller
than the largest supercurrent circulating around vortices, as discussed below.

\begin{figure}[tb]
\includegraphics[width=\columnwidth]{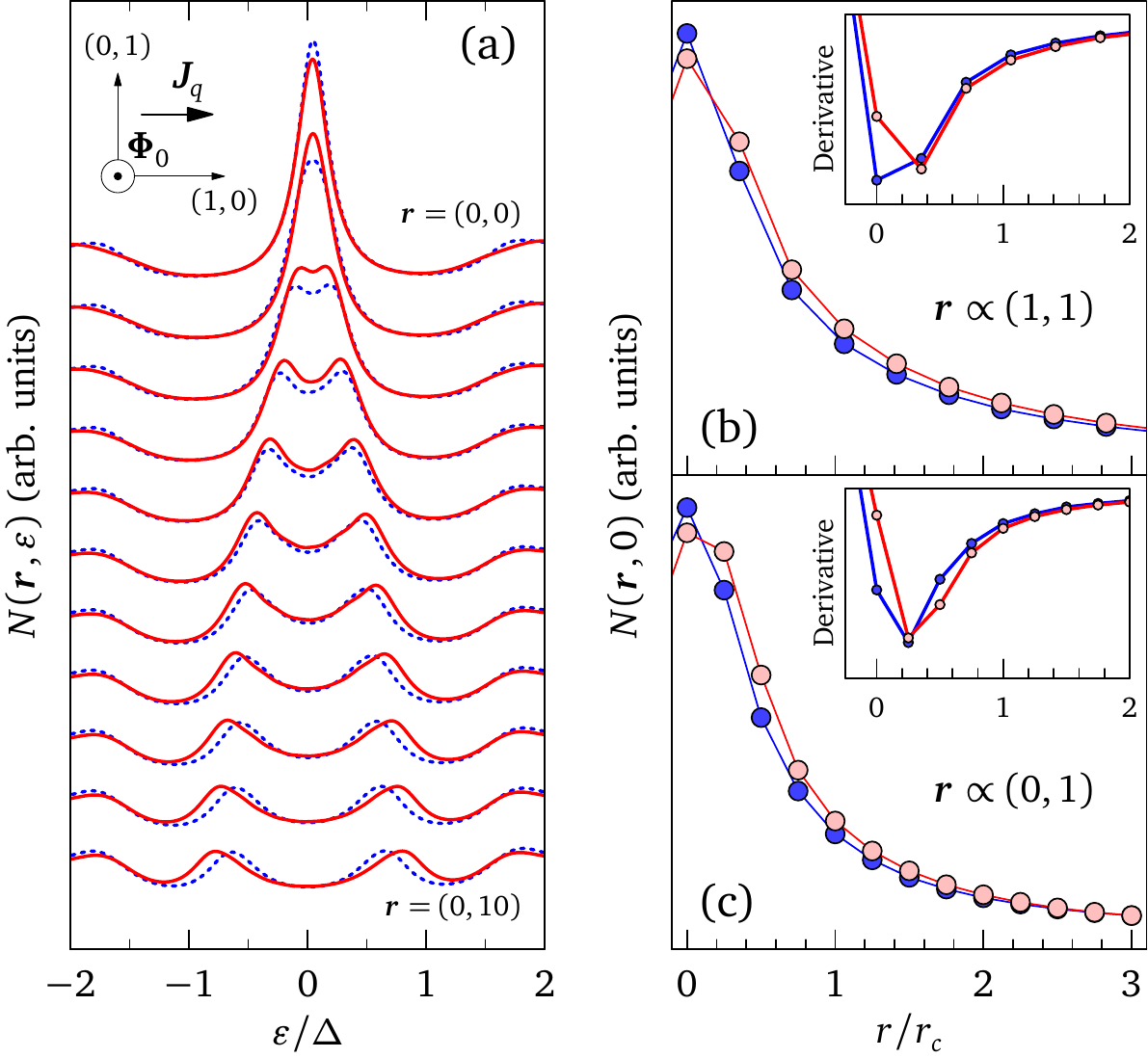}
\caption{\label{fig:fig3}
(a) LDOS along the $(0,1)$ direction for an isolated vortex at position $(0,0)$
with $t=10$ and $r_c=4$. The dotted lines are for $q=0$ and the solid lines for
$q=0.02$. The inset shows the direction of the applied current; $\vec{\Phi}_0$
is the vortex magnetic flux vector. The zero-energy LDOS along the diagonal and
along the $y$ axis are shown in (b) and (c), respectively, without (full
symbols) and with (empty symbols) the applied current. The insets show
$dN(r,0)/dr$ obtained by numerical differentiation.
}
\end{figure}

We now turn to the case of an isolated vortex in a uniform applied current. For
the order parameter we assume the form
$\Psi(\vec{r})=\Delta\tanh(r/r_c)e^{-i(\vartheta+qx)}$, where
$\vec{r}=(x,y)=r(\cos\vartheta,\sin\vartheta)$, the origin being at the center
of the $M\times M$ mesh. The vortex-core radius $r_c$ is estimated as
$r_c\sim\xi$, with $k_{\mathrm{F}}\xi=(4/\pi)(t/\Delta)$, and
$k_{\mathrm{F}}\approx\pi/(2a)$. We thus obtain $r_c/a=(2/\pi)^2(t/\Delta)$. The
ratio of the applied and vortex currents is $J_q/J_{\chi}=qr$, and the vortex
current is largest at $r=r_c$. Therefore, if $qr_c\ll1$, the applied current is
much smaller than the maximum vortex-induced supercurrent. For $t=2.5$ and
$t=10$, we have $r_c\approx 1$ and $r_c\approx 4$, respectively, such that the
condition is satisfied with the respective values $q=0.05$ and $q=0.02$. The
model assumes that the vortex is pinned without being actually close to a
pinning center. This is appropriate in a regime of collective pinning, as in the
NbSe$_2$ experiments.

For $q=0$, the calculated vortex LDOS shown in Fig.~\ref{fig:fig3}(a) exhibits
the well-known structures common to BCS $s$-wave vortices \cite{Caroli-1964,
Bardeen-1969, Shore-1989, Hayashi-1998, Gygi-1990a, *Gygi-1991}: a low-energy
peak at the vortex center, which splits with increasing distance from the
center. In the presence of the current, the central peak is reduced, while the
zero-energy LDOS increases with respect to the zero-current case when moving
outside the core. The trend is opposite slightly below the gap edges: the LDOS
is enhanced in the core and reduced outside the core. Figures~\ref{fig:fig3}(b)
and \ref{fig:fig3}(c) compare the zero-energy LDOS and its numerical derivative,
with and without the current, on the lines going from the center of the vortex
along the $(1,1)$ and $(0,1)$ directions, respectively. These results show
striking similarities with the experiment \cite{Maldonado-2013}, in particular,
an apparent increase of the vortex-core size revealed by a displacement of the
minimum in the LDOS derivative. The figure also suggests that the energy
separation between the core states is reduced by the uniform current for
$r\lesssim r_c$, and increased for $r>r_c$. The same behavior is observed for
$t=2.5$, which excludes a finite-size effect. This phenomenon is related to the
polarization of the vortex-core states, as discussed further below.

\begin{figure}[tb]
\includegraphics[width=\columnwidth]{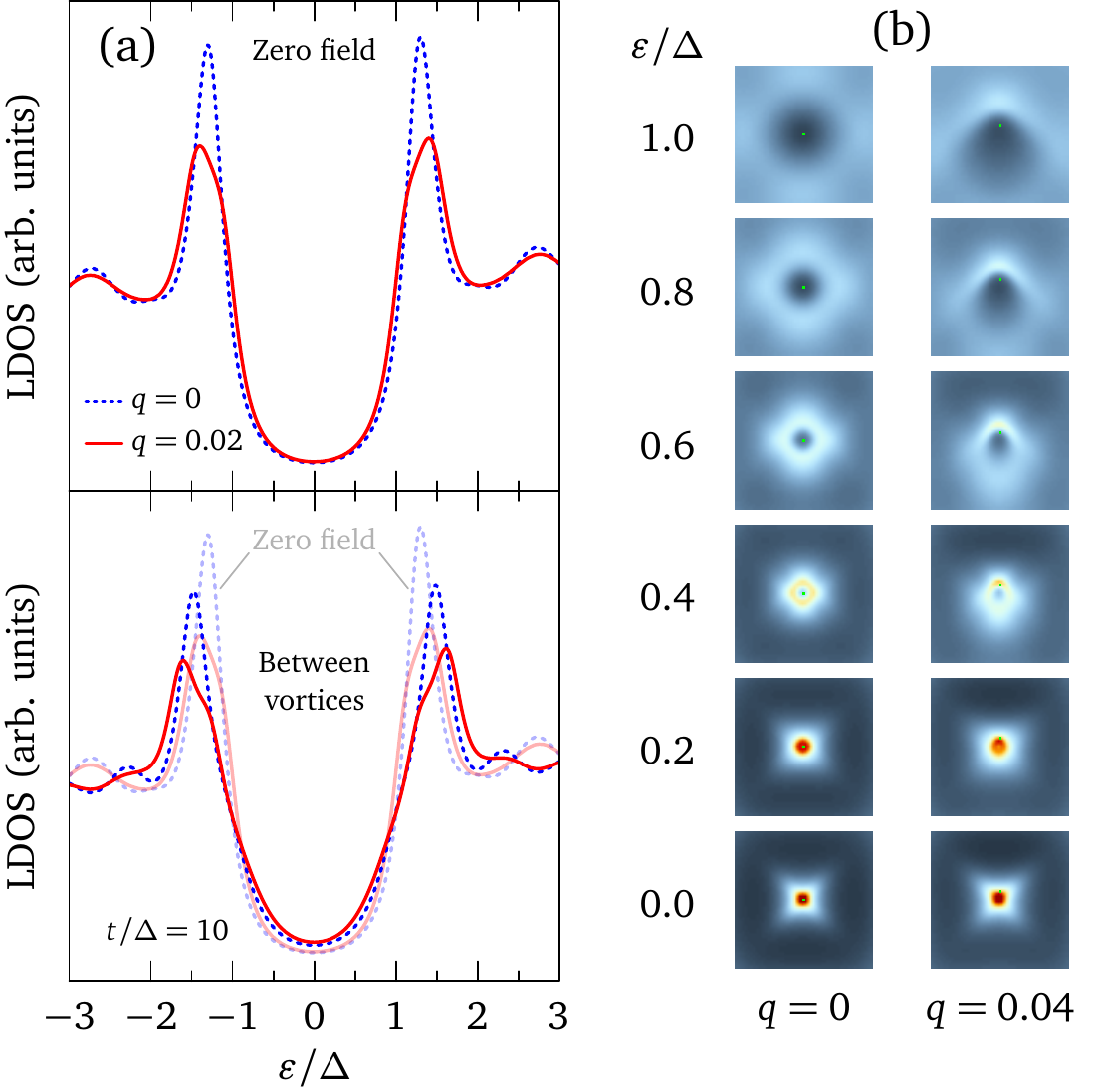}
\caption{\label{fig:fig4}
(a) Zero-field LDOS at the center of the $71\times71$ mesh (top panel) and LDOS
in a triangular vortex lattice at equal distance from three neighboring vortices
(bottom panel: the data of the top panel is repeated for comparison). The dotted
lines are for $q=0$ and the solid lines for $q=0.02$. (b) LDOS in real space for
a vortex core in a vortex lattice at various energies without (left) and with
(right) an applied current. The region shown is the $51\times51$ central part of
the $71\times 71$ mesh. The color scale is the same in all images. The green dot
is the point where $|\Psi(\vec{r})|=0$.
}
\end{figure}

If the vortex belongs to a vortex lattice, we found that the core spectra are
slightly broadened with respect to those in Fig.~\ref{fig:fig3} but that the
general trends remain unchanged. We considered a triangular vortex lattice with
a nearest-neighbor vortex distance of $50$. With this value, the LDOS far from
vortices differs from the zero-field LDOS, as shown in Fig.~\ref{fig:fig4}(a):
there are more states in the gap at finite field, the peaks are reduced, and the
gap appears slightly larger, in good qualitative agreement with the NbSe$_2$
data \cite{Maldonado-2013}. Figure~\ref{fig:fig4}(b) compares the vortex-core
LDOS in a vortex lattice with and without the applied current. The
current-induced expansion of the core size can be distinguished at low energy
(we used here a larger current $q=0.04$ in order to emphasize this). Note that
the LDOS has an energy-dependent fourfold anisotropy due to the underlying
square symmetry of the model \cite{Zhu-1995}. At higher magnetic fields
(intervortex distance $\lesssim 30$), a sixfold anisotropy develops due to the
vortex lattice.

The images in Fig.~\ref{fig:fig4}(b) for $q=0.04$ show a systematic deformation
in the direction $(0,-1)$, which is the direction of the force
$\vec{J}_q\times\vec{\Phi}_0$ (downwards in the figures). At zero energy, the
LDOS peak does not coincide with the point where $|\Psi(\vec{r})|=0$, but is
shifted by the applied current in the direction of the force. In
Figs.~\ref{fig:fig3} and \ref{fig:fig4}(b), this has been corrected by
displacing the origin in the direction $(0,-1)$ by two and three lattice
spacings, respectively, in the finite-current data. With increasing energy, the
center of gravity of the vortex bound states moves further in the direction of
the force. Thus the whole electronic structure of the vortex is bent by the
applied current. The spatial separation between the zero of $|\Psi(\vec{r})|$
and the center of the bound states is another illustration of the key role
played by the order-parameter \emph{phase} in the formation of the vortex
states, and the marginal relevance of its modulus \cite{Berthod-2005}. A
displacement of the LDOS peak with respect to the phase singularity point was
also found in vortex-antivortex pairs \cite{Melnikov-2009}. Because
$\Psi(\vec{r})$ is artificially pinned in our non-self-consistent calculations
and the high-energy states must remain orthogonal to the low-lying ones, the
wave functions sharpen on one side of the vortex and extend on the other side,
leading to the characteristic polarization seen in Fig.~\ref{fig:fig4}(b). This
polarization explains the shift of the core-state peaks to higher energies in
Fig.~\ref{fig:fig3}(a). In the direction $(0,1)$, the bound states pile up more
densely in real space, and the core-state peaks disperse more rapidly with
distance. The opposite behavior occurs in the direction $(0,-1)$, where the
core-state peaks are shifted to \emph{lower} energies (not shown in the figure)
with respect to the zero-current LDOS. No energy shift, but a slight deformation
of the peaks, is observed in the $x$ direction parallel to the current.
Observing the polarization of the LDOS in the direction normal to the current is
an interesting challenge for future STM experiments.

The origin of the force acting on vortices in the presence of a supercurrent has
been discussed by many authors \cite{Kopnin-1976, Volovik-1993a, Stone-1996,
Chen-1998, Narayan-2003, Melnikov-2011}. Our calculations show that the
superflow transfers momentum into the bound states, resulting in a polarization
of the wave functions if the vortex is pinned. On one side of the vortex, the
bound states are more localized, because the applied supercurrent is contrary to
the vortex supercurrent and the superfluid velocity is reduced. On the other
side, the two supercurrents add up and the wave function is more extended. This
effect may be considered to have a magnetic origin, the vector potential of the
applied current changes the phase relation between the electron and hole parts
of the Bogoliubov excitations in the vortex, but is obviously different from the
electromagnetic interaction between the applied supercurrent and the magnetic
flux carried by the vortex. The polarizability of the vortex-core states has not
been considered so far in the study of the interaction between currents and
vortices, and between different vortices. A microscopic calculation of the
vortex energy in a uniform applied current would be a first step in this
direction. This is not an easy task, however, because a self-consistent
determination of the fields and currents is required for a precise comparison of
the various forces.

\end{document}